\documentclass[fleqn,twoside]{article}
\usepackage{espcrc2,colordvi,psfig}

\title{Free energy and $\theta$ dependence of SU($N$) gauge 
theories\thanks{Talk presented by H. Panagopoulos}}

\author{L. Del Debbio\address{Dipartimento di Fisica, Universit\`a di Pisa,
and INFN, I-56127 Pisa, Italy},
        H. Panagopoulos\address{Department of Physics, University of
        Cyprus, Nicosia CY-1678, Cyprus},
        E. Vicari$^a$}

\begin{document}

\begin{abstract}
We study the dependence of the free energy on the CP violating angle
$\theta$, in four-dimensional
SU($N$) gauge theories with $N\geq 3$, and 
in the large-$N$ limit.
Using the Wilson lattice formulation for numerical simulations, 
we compute the first few terms of the 
expansion of the ground-state energy $F(\theta)$ around $\theta=0$,
$F(\theta)-F(0) = A_2 \,\theta^2 ( 1 + b_2 \theta^2 + ...)$.
Our results support Witten's conjecture:
$F(\theta)-F(0) = {\cal A}\,\theta^2 + O(1/N)$
for $\theta < \pi$. 
We  verify that the topological susceptibility 
has a nonzero large-$N$ limit $\chi_\infty=2 {\cal A}$ with corrections of
$O(1/N^2)$, in substantial agreement with the Witten-Veneziano formula
which relates $\chi_\infty$ to the $\eta^\prime$ mass. Furthermore,
higher order terms in $\theta$ are suppressed; in particular, the
$O(\theta^4)$ term $b_2$ (related to the $\eta^\prime - \eta^\prime$
elastic scattering amplitude) turns out to be quite small:
$b_2=-0.023(7)$ for $N=3$, and its absolute value decreases with
increasing $N$, consistently with the expectation $b_2=O(1/N^2)$.

\end{abstract}

\maketitle
\section{\Blue{INTRODUCTION}}

Interest in $SU(N)$ gauge theories has recently been revived by 
studies in the context of M-theory, AdS/CFT correspondence, and 
${\cal N}=2$ SUSY broken to ${\cal N}=1$. A number of
predictions have been put forth, and we have set out to check these
predictions by numerical simulation.

One aspect of these studies regards the spectrum of confining strings in
four-dimensional SU($N$) gauge theories. A considerable amount of
numerical work has been done on this subject (see~\cite{LT-01,DPRV}
and references therein). In particular, in Ref.~\cite{DPRV}
we have presented computations of the string tensions $\sigma_k$,
related to sources with $Z_N$ charge $k$, for SU(4) and SU(6) gauge
theories. Our results are
consistent with the sine formula
$\sigma_k/\sigma_1 = \sin (k\,\pi/N) / \sin (\pi/N)$, with an accuracy of
approximately 1\% [2\%] for SU(4) [SU(6)]; this formula emerges in various
realizations of supersymmetric SU($N$) gauge theories, and in the
context of M-theory. On the other hand, our results show
deviations from Casimir scaling $\sigma_k/\sigma = {k (N-k)/(N-1)}$,
as might be expected from the short distance behaviour of the
potential, or from a strong-coupling limit.

In the present work, we study the $\theta$ dependence of SU($N$) gauge
theories, for $N\geq 3$ and in the large-$N$ limit, by
numerical simulation; for a longer write-up of our work, see~\cite{DPV}.
The angle $\theta$ appears in the Euclidean Lagrangian as:
\begin{equation}
{\cal L}_\theta  = {1\over 4} F_{\mu\nu}^a(x)F_{\mu\nu}^a(x)
- i \theta q(x)
\label{lagrangian}
\end{equation}
where $q(x)$ is the topological charge density
\begin{equation}
q(x) = {g^2\over 64\pi^2} \epsilon_{\mu\nu\rho\sigma}
F_{\mu\nu}^a(x)F_{\rho\sigma}^a(x).
\end{equation}

The partition function reads:
\begin{equation}
Z(\theta) {=} \int\! [dA] \exp ( {-} \int\! d^4 x {\cal L}_\theta )
\equiv \exp[ - V F(\theta) ]
\end{equation}
where $F(\theta)$ is the free (ground state) energy.
Witten has argued \cite{Witten-80} that in the large-$N$ limit
$F(\theta)$ is a multibranched function of the type
\begin{equation}
F(\theta) = N^2 {\rm min}_k\, H[ (\theta+2\pi k)/ N]
\label{conj1}
\end{equation}
The conjecture was refined \cite{Witten-98} and simplified:
\begin{equation}
\Delta F(\theta) = {\cal A} \, {\rm min}_k \, (\theta+2\pi k)^2 + O\left(
1/N\right).
\label{conj2}
\end{equation} 
($\Delta F(\theta)\equiv F(\theta)-F(0)$).
In particular, for $\theta<\pi$, 
\begin{equation}
\Delta F(\theta) = {\cal A} \, \theta^2  + O\left( 1/N\right).
\label{conj2b}
\end{equation} 
This conjecture has been supported by gauge/string duality arguments~\cite{Witten-98}.

Monte Carlo studies of the $\theta$-dependence are made 
very difficult by the complex nature of the $\theta$ term.
In fact $\theta\ne 0$ cannot be directly simulated on the lattice.
Here we restrict ourselves to 
relatively small $\theta$ values, 
where one may expand:
\begin{equation}
\Delta F(\theta) = A_2 \theta^2 + A_4 \theta^4 + \ldots .
\label{expf}
\end{equation}
$A_2$ gives the topological susceptibility, i.e. 
\begin{equation}
A_{2} =  \chi/2, \qquad \chi 
=  \langle Q^2 \rangle_{\theta=0}/V,
\end{equation}
where $Q$ is the topological charge $Q=\int d^4 x\, q(x)$.
Higher order coefficients in Eq.~(\ref{expf})
can be related to higher moments of the probability distribution
$p(Q)$ of the topological charge.
For instance,
\begin{equation}
A_4 = - {1\over 24 V} \left[ 
\langle Q^4 \rangle - 3 \langle Q^2 \rangle ^2 \right]_{\theta=0}. \label{chi4}
\end{equation}
A convenient parameterization of the free energy using dimensionless
scaling coefficients is:
\begin{equation}
\sigma^{-2}\,\Delta
F(\theta)={1\over 2} C \,\theta^2 (1 + b_{2} \theta^2 + b_4 \theta^4 +
...)
\end{equation} 
($\sigma$: string tension, $C={\chi / \sigma^2}$, $b_2 = {A_4 / A_2}$).

Witten's conjecture implies that
$C$ has  a finite nonzero large-$N$ limit $C_{\infty}$ 
and $b_{2i}=O(1/N^2)$.
In particular one expects that $b_2$ is small, and that it should
rapidly decrease with increasing $N$.
A nonzero value of $\chi_{\infty}$ is essential
to resolve the U(1)$_A$ problem, and is related 
to the $\eta'$ mass \cite{Witten-79}:
\begin{equation}
\chi_\infty = 2 {\cal A} = {f_\pi^2 m_{\eta'}^2\over 4 N_f} + O(1/N).
\label{wittenformula}
\end{equation}

\section{\Blue{CALCULATION AND RESULTS}}
We present results for four-dimensional
SU($N$) gauge theories with $N=3,4,6$.
In a nutshell, we: Determine the ratio $C=\chi/\sigma^2$;
verify the expected behavior $C \sim A+B/N^2$, determining the large-$N$ limit by a fit;
obtain a rather accurate estimate of $C_\infty$;
estimate the values of $b_2$, which are found to be very small 
and rapidly decreasing with $N$, in support of Witten's conjecture.

\bigskip\noindent
\underbar{\bf Simulation Details:} $\bullet$ We use the standard Wilson gauge action
and the 
Cabibbo-Marinari updating algorithm 
(one SU(N) update $\to N(N-1)/2$  SU(2) subgroup updates).
$\bullet$ Microcanonical over-relaxation and heat-bath updates are used in a 4:1 ratio.
$\bullet$ Values for 
$\beta$ are chosen in the weak-coupling,
(beyond the 1st order phase transition ($N=6$), beyond crossover (peak of specific
heat $N=3,4$)).
$\bullet$ Topological quantities are computed via cooling.
We measure $Q$ typically every 100 sweeps. $\bullet$ We have employed
different lattice sizes, so that finite size effects 
are under control. Tables of our numerical results can be found in
Ref.~\cite{DPV}.

\vskip 1cm 

\centerline{\psfig{width=7.2truecm,file=tau.eps}}

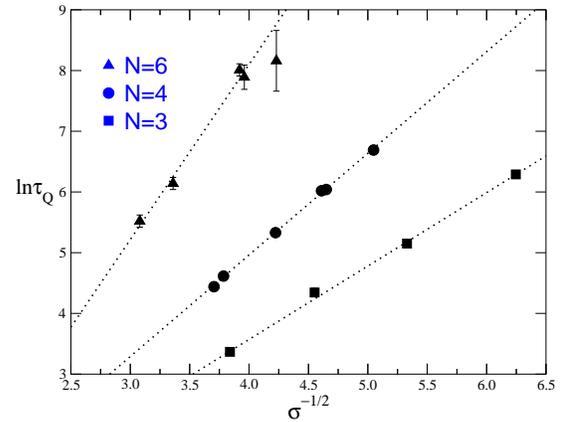
\noindent{\small {\tt FIGURE 1. } Autocorrelation time \Red{$\tau_Q$} versus
\Red{$\xi_\sigma\equiv \sigma^{-1/2}$}. Dotted lines show the linear fits.}

\medskip
A severe form of critical slowing down is seen in the measurement of $Q$,
at large values of $N$.
Our results (Fig. 1) suggest an exponential behavior:
\begin{equation}
\ln \tau_Q \approx N \left( e_N \xi_\sigma + c_N \right)
\end{equation}
($\xi_\sigma\equiv \sigma^{-1/2}$,
and $e_N$, $c_N$ are constants with only weak $N$-dependence).
Thus, $\ln \tau_Q \propto N$ (at fixed $\xi_\sigma$).
This phenomenon is possibly due to the free-energy barriers present. There are
no a priori arguments in favor of this
behavior; it simply arises naturally from fits, unlike a power law.
No similar effect is observed 
in plaquette or Polyakov line correlations; this 
suggests a decoupling between topological modes and nontopological ones,
such as those determining confining properties.  

In Fig. 2, the ratio $C$ is plotted versus $\sigma$, 
to evidentiate possible scaling corrections (expected: ${\cal
O}(a^2)$, logs).
Corrections are indeed observed for $N=3$
and $N=4$, and we must allow for them in a fit, for a continuum extrapolation.
Assuming a linear behavior: $C= C_{\rm cont} + u \sigma$,
to take into account the leading
${\cal O}(a^2)$  correction, we obtain:
\begin{eqnarray}
C_{\rm cont} = 0.0282(12), u=0.077(24) &(N{=}3) \\
C_{\rm cont} = 0.0257(10), u=0.049(15) &(N{=}4). 
\end{eqnarray}
For $N=6$, there is no evidence of scaling corrections; we find: $C_{\rm cont} =
0.0236(10)$ for $N=6$.

\vskip 1cm 

\centerline{\psfig{width=7.2truecm,file=chi2.eps}}

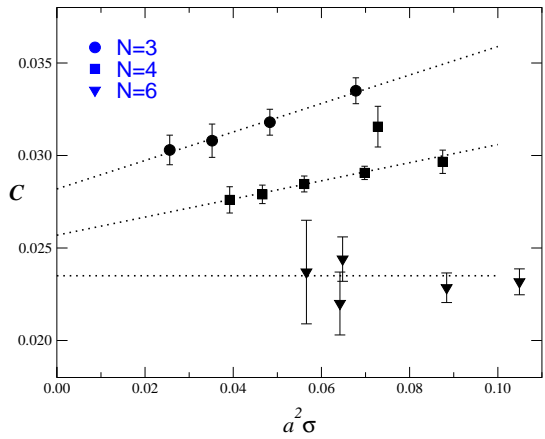
\noindent{\small {\tt FIGURE 2. } The scaling ratio
  \Red{$C=\chi/\sigma^2$} versus \Red{$\sigma$}. Dotted lines show the linear fits.}

\medskip
In Fig. 3, our results for $C_{\rm cont}$ are plotted versus $1/N^2$
(the expected order of $1/N$ corrections).
They are clearly consistent with a behavior of
the type $C_{\rm cont}(N) = C_\infty + B/ N^2.$
Assuming such a behavior, a fit to $N=3,4,6$ results gives
\begin{equation}
C_\infty = 0.0221(13), \qquad B=0.055(19),
\end{equation}
This is a rather accurate estimate of
large-$N$ limit of $C$, and it is clearly nonzero.

To compare with literature, we set a standard value: $\sqrt{\sigma}=440 \,
{\rm MeV}$, for $N \geq 3$. 
We obtain: 
$\chi_\infty^{1/4}=170(3) \, {\rm MeV}$, and $\chi^{1/4}=180(2) \,{\rm
MeV}$ for $N=3$. 
From Eq.~(\ref{wittenformula}), with actual values of $f_\pi$,
$m_{\eta '}$, and $N_f=3$, one has $191 \, {\rm MeV}$.
Using Veneziano's improvement: $m_{\eta'}^2 \rightarrow
m_{\eta'}^2 + m_\eta^2 - 2 m_K^2$, one obtains $180 \, {\rm MeV}$.

We now turn to $b_2$:
The cancellations in its definition necessitate very high statistics.
We have obtained sufficiently accurate  results for $N=3,4$, showing
scaling.
A weighted average, with conservative errors, yields:
\begin{equation}
b_2 {=} - 0.023(7) (N{=}3),\  
b_2 {=} - 0.013(7) (N{=}4).
\end{equation}

Reducing statistical error on $b_2$ in the $N=6$ case is extremely
difficult. We can only give a rough estimate $b_2=-0.01(2)$
($\gamma=0.348$) in this case, showing that $b_2$ is very small.

\vskip 1cm

\centerline{\psfig{width=7.2truecm,file=summary.eps}} 

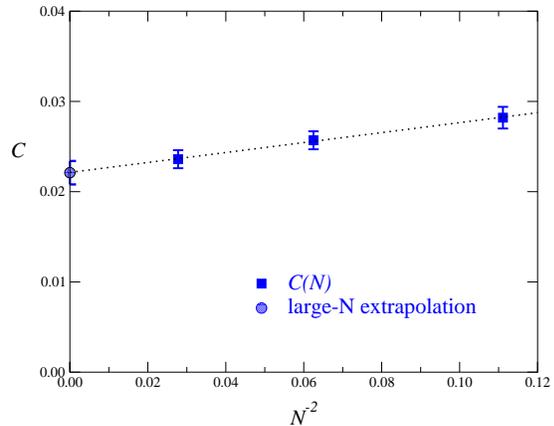
\noindent{\small {\tt FIGURE 3. } The scaling ratio
  \Red{$C=\chi/\sigma^2$} versus \Red{$1/N^2$}. The line is the
  linear fit to \CarnationPink{$A+B/N^2$}.} 

\medskip
An alternative to cooling is based on the
index theorem, using, e.g., the overlap formulation
\cite{Neuberger-98}. The rigorous properties of 
such an approach
would make up for a useful consistency check.

In conclusion, we stress that further investigation of
other observables in $SU(N)$ theories is called
for, to test predictions from:
SUSY $SU(N)$, M-theory and gauge/string duality.

\end{document}